\documentclass[a4paper,twocolumn,11pt]{quantumarticle}
\pdfoutput=1
\usepackage[utf8]{inputenc}
\usepackage[english]{babel}
\usepackage[T1]{fontenc}
\usepackage{amsmath}
\usepackage{hyperref}
\usepackage{tabularx}
\usepackage{graphicx}
\usepackage[T1]{fontenc}
\usepackage{xcolor}
\usepackage{amsmath}
\usepackage{multirow}

\graphicspath{{FIG/},{POK/}}

\newcommand{\FEFW}{FEF$_{\rm w}$}

\begin{document}

\title{Machine classification of quantum correlations for entanglement distribution networks}

\author{Jan Soubusta} \email{jan.soubusta@upol.cz}
\affiliation{Institute of Physics of the Academy of Sciences of the Czech Republic, Joint Laboratory of Optics of Palacký University and Institute of Physics AS CR, 17. listopadu 50a, 772 07 Olomouc, Czech Republic}
\affiliation{Palacký University in Olomouc, Faculty of Science, Joint Laboratory of Optics of Palacký University and Institute of Physics AS CR, 17. listopadu 12, 771 46 Olomouc, Czech Republic}

\author{Antonín Černoch} 
\affiliation{Institute of Physics of the Academy of Sciences of the Czech Republic, Joint Laboratory of Optics of Palacký University and Institute of Physics AS CR, 17. listopadu 50a, 772 07 Olomouc, Czech Republic}

\author{Karel Lemr} \email{k.lemr@upol.cz}
\affiliation{Palacký University in Olomouc, Faculty of Science, Joint Laboratory of Optics of Palacký University and Institute of Physics AS CR, 17. listopadu 12, 771 46 Olomouc, Czech Republic}

\begin{abstract}
The paper suggest employing machine learning for resource-efficient classification of quantum 
correlations in entanglement distribution networks.
Specifically, artificial neural networks (ANN) are utilized to classify quantum 
correlations based on collective measurements conducted in the geometry of entanglement swapping. 
ANNs are trained to categorize two-qubit quantum states into 
five mutually exclusive classes depending on the strength of quantum correlations 
exhibited by the states. The precision and recall of the ANN models are analyzed as functions of the quantum resources consumed, i.e. the number of collective measurements performed.
\end{abstract}



\section{Introduction}

Quantum correlations naturally emerge as a consequence of the principle of superposition
\cite{Dirac_1948,Nielsen:QIP} 
occasionally leading to observations that are inconsistent with 
the rules of classical physics \cite{Bell64,CHSH69,Aspect81,Aspect82}. 
Beyond their profound philosophical implications, quantum correlations serve as 
the bedrock for numerous cutting-edge quantum technologies 
\cite{jaeger:QIP,Zeilinger:QT,Pan:swapp,Ekert91}.
They enable parallelism yielding quantum-computational speed-up 
\cite{Quanta38,Layden2023,Abbott2010} 
and contribute to the superior precision in quantum metrology 
\cite{RevModPhys.89.035002,RevModPhys.90.035005,RevModPhys.92.015004}.
Quantum correlations play an indispensable role in quantum cryptography as well, 
where it has been demonstrated that nonlocality constitutes security guarantees even 
when using untrusted devices \cite{Tan2022improveddiqkd,Zhen2023}. 
Meanwhile, steerable quantum states suffice to safely operate in a single-sided 
untrusted regime \cite{Xin:20}.
The distribution of quantum states manifesting strong nonclassical correlations 
facilitates quantum communications in general. Overcoming losses that would otherwise 
impede long-distance quantum communications is achievable through the utilization 
of quantum repeaters \cite{Briegel:repeater}, relays \cite{Jacobs:relay}, 
or teleportation-based quantum networks \cite{barasinski:Qinternet}, 
all employing entanglement distribution and the entanglement swapping protocol.

\begin{figure}
\centering
 \includegraphics[width=\linewidth]{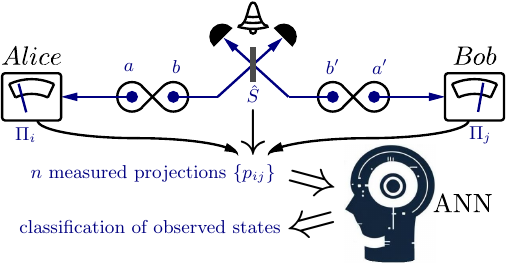}
 \caption{\label{fig_schema} Conceptual diagram of quantum state classification
 from collective measurements using ANN. Two copies of 
 a two-qubit state are subject to collective measurements yielding a vector of 
 measurement outcomes $\{p_{ij}\}$. This feature vector of length $n$ is processed 
 by the ANN that predicts the state's class.}
\end{figure}

Evaluation of the quality of distributed quantum states becomes imperative
in practical implementations of these communications networks. 
Furthermore, this assessment must be both highly accurate and rapid to avoid excessive 
consumption of expensive quantum resources. An elegant solution is found in the 
framework of collective measurements, applicable in the same geometric setup as 
entanglement swapping (refer to the conceptual diagram in Fig.~\ref{fig_schema})
\cite{Rudnicki_2011,Rudnicki_2012,Bartkiewicz.PRA.95,
Bartkiewicz_2017,Bartkiewicz_2018,Travnicek_2018}.
While some quantum correlations can be unveiled through analytical calculations 
based on the outcomes of measurements performed in this setting, other, such as 
the negativity, are inaccessible this way \cite{Bartkiewicz.PRA.95}. 
A remedy lies in harnessing 
the predictive capability of artificial neural networks (ANN), renowned for 
their ability to predict quantum state properties even from incomplete measurements
\cite{Tomamichel2011,Jezek2023,Roik_2021,Roik_2022,Greenwood2023}.

In this paper, we analyze the accuracy of detecting quantum state correlations through 
ANNs using collective measurements acquired in the entanglement swapping layout. 
Additionally, we evaluate the performance of ANNs under gradually reduced set 
of measurements determining the trade-off between accuracy and resource consumption
\cite{MML_Deisenroth}. We develop and test a method for optimization of this 
trade-off by employing the inner symmetry of the problem in two-qubit Hilbert space.
Our focus is directed towards five important categories of quantum correlations and 
classification of quantum states into five corresponding and mutually exclusive groups 
based on the \emph{strongest} quantum correlation categorized states can produce: 
Bell nonlocal states \cite{Bell64,CHSH69,Aspect81}; 
steerable \cite{Wiseman2007,Fan2021}, but Bell local states; 
unsteerable states with fully entangled fraction (FEF) above the threshold of 
pure separable state (FEF $>\frac{1}{d}$, $d=2$ for qubits)
\cite{Bennett1996,Grondalski2002,Zhao_2010,Bartkiewicz_2017,Patro2022}; 
entangled states not belonging to the previous class, 
i.e. FEF $\leq\frac{1}{d}$; 
and finally separable states (see summary in Table~\ref{Tab_schody}).

To enhance readability of the paper, the term \emph{entangled 
states} refers to states that do not posses FEF above 
$\frac{1}{d}$. Similarly, the term \emph{steerable states} designate 
states that are steerable, but not Bell non-local. The adoption of these five categories 
is justified by potential applications of nonlocal and steerable states in 
device-independent quantum cryptography and the entanglement as a resource unique to 
quantum technologies. Formal definitions are introduced in the next section.

\begin{table}
\caption{\label{Tab_schody} Hierarchy of quantum correlations used for 
         state classification. Quantities used in the class definitions are 
         formally introduced in Sec.~\ref{sec:classes}.} 
%
\begin{tabularx}{\linewidth}{c X c}
\hline\hline
Class & Description & Definition \\
\hline 
sep   & separable    & $N = 0$\\
ent   & entangled    & $N > 0$ and \FEFW = 0\\
FEF   & fully ent. frac.   & \FEFW > 0 and $S_3 = 0$\\
steer & steerable   & $S_3 > 0$ and $B = 0$\\
Bell  & Bell non-local   & $B > 0$\\
\hline\hline
\end{tabularx}
\end{table}


\section{Quantum states generation and classification}
\label{sec:classes}

To begin, we generated a collection of 160 million random two-qubit quantum 
states using the method outlined in \cite{Pozniak_1998,Maziero2015}. 
Following this, we employed analytical formulas based on the density matrix of 
each state to classify them \cite{Abo2023}. 
The true labels $c$ acquired through this 
classification are utilized to subsequently train and test the ANN models later on. 
Negativity \cite{Zyczkowski1998} 
was obtained from the eigenvalues of the partially transposed 
density matrix $\lbrace \lambda_i \rbrace = \mathrm{eigs} (\hat{\rho}^\Gamma)$
\begin{equation}\label{EQ_Negativity}
  N(\hat{\rho}) = \sum_i |\lambda_i| - \lambda_i.
\end{equation}
The remaining quantities were calculated using the correlation matrix $R = T^T T$, where
\begin{equation}\label{EQ_Tmatrix}
  T_{ij}(\hat{\rho}) = \mathrm{Tr}\left[\hat{\rho} (\sigma_i\otimes\sigma_j)\right]
\end{equation}
and $\sigma_i$ for $i\in[1,3]$ are Pauli matrices.
According to its usual definition, FEF is expressed as the overlap between the tested 
state $\hat{\rho}$ and a maximally entangled state $|\Phi\rangle$ maximized over all 
maximally entangled states 
\cite{Grondalski2002},
$$
  \mathrm{FEF}(\hat{\rho}) \equiv  \underset{|\Phi\rangle}{\max}
  \left\{\langle \Phi| \hat{\rho} |\Phi \rangle\right\}.
$$
Instead of this standard definition, we use a more convenient fully entangled 
fraction witness, 
\begin{equation}\label{EQ_FEF}
  \mathrm{FEF_w}(\hat{\rho}) = \frac{1}{2} \mathrm{max}
  \left\{ 0, \mathrm{Tr}\sqrt{R}-1 \right\},
\end{equation}
which relates to FEF in a way that \FEFW $> 0$ iff FEF $>{1\over d}$.
Similarly, the Costa–Angelo 3-measurement steering \cite{Costa2016} is 
calculated as
\begin{equation}\label{EQ_S3}
  S_3(\hat{\rho}) = \sqrt{\frac{1}{2}\mathrm{max}\left\{ 0,\mathrm{Tr}R-1\right \}}
\end{equation}
and the Bell nonlocality \cite{Horodecki1995,Adam2004,Yang2021},
\begin{equation}\label{EQ_Bell}
  B(\hat{\rho}) = \sqrt{\mathrm{max}\lbrace 0, \mathrm{Tr}R 
                     -\mathrm{min}\left[\mathrm{eig}(R)\right] -1\rbrace}.
\end{equation}
For the labeling of generated states using the 
above-defined quantities see Tab.~\ref{Tab_schody}.

The correlation matrix $R$ can be efficiently obtained from a set of collective
measurements implemented on two copies of the investigated state \cite{Horst_2013}
\begin{equation}\label{eq:R_from_collectives}
  R_{ij} = \mathrm{Tr}\left[\hat{\rho}_{ab}\otimes\hat{\rho}_{a'b'} \hat{S}_{bb'}
  \left(\sigma_i\otimes\sigma_j\right)_{aa'}\right],
\end{equation}
where $\hat{S} = 1 - 4|\Psi^-\rangle\langle\Psi^-|$ and subscripts 
$\lbrace a,b,a',b'\rbrace$ denote subsystems of the first and second copy 
of the investigated state. $|\Psi^-\rangle$ is the maximally entangled singlet 
Bell state.  

Importantly, negativity can not be calculated from the correlation matrix $R$ as 
it would require a more complex collective measurement strategy \cite{Bartkiewicz.PRA.95}. 
Eq. (\ref{eq:R_from_collectives}) stipulates that the correlation matrix $R$ 
can be obtained by applying 16 measurement configurations each corresponding 
to a combination of two {\it von Neumann projectors} applied to subsystems 
$a$ and $a'$ (see conceptual scheme in Fig.~\ref{fig_schema})
while subsystems $b$ and $b'$ are always projected onto a singlet Bell state.
For a detailed reasoning see subsection~\ref{App:minimal} of the Appendix. 

\begin{figure}[h]
\centering
 a)\includegraphics[width=.4\columnwidth]{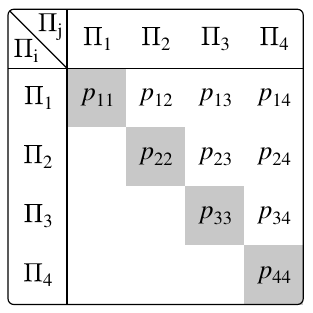}
 \hfill 
 b)\includegraphics[width=.4\columnwidth]{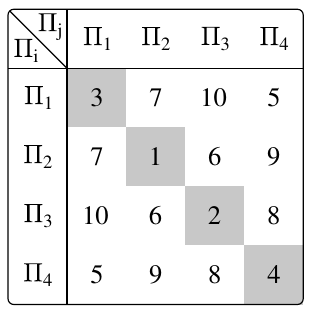}
 \caption{\label{Fig_Features} a) Labeling of collective two-qubit 
          projections by indexes of $p_{ij}$ defined in Eq. (\ref{EQ_pij}). 
          b) {Feature ordering used in the higher-number projection strategy.}
          Reduction is performed in the descending order of the values.}
\end{figure}

One can further reduce the number of measurements by replacing the projectors based 
on Pauli matrices by the so-called minimal basis set \cite{Rehacek2004}. The minimal 
basis set of projectors $\Pi_i$ for $i\in[0,3]$ is related to the Pauli matrices 
by a simple matrix transform $\Pi_i = \sum_j M_{ij} \sigma_j$ 
(assuming $\sigma_0 = 1$). For the details on the minimal basis projectors 
$\Pi_i$ and the matrix $M$, see subsection \ref{App:minimal} in the Appendix. 
Replacing Pauli matrices in Eq. 
(\ref{eq:R_from_collectives}) by $\sum_j M_{ij} \Pi_j$ and considering symmetry of
$$
  p_{ij} = p_{ji}
$$
for
\begin{equation}\label{EQ_pij}
  p_{ij} = \mathrm{Tr}\left[\hat{\rho}_{ab}\otimes\hat{\rho}_{a'b'} \hat{S}_{bb'}
  \left(\Pi_i\otimes\Pi_j\right)_{aa'}\right]
\end{equation}
decreases the number of measurement configurations needed to establish the 
correlation matrix $R$ to only 10 measurements.
Graphical representation of the symmetry of measured quantities $p_{ij}$
as well as their unique numbering from 1 to 10 is presented in 
Fig.~\ref{Fig_Features}.

\section{Implementation of the ANN models}

We have implemented a series of ANN models to evaluate the performance of 
ANNs in classification of quantum states according to the hierarchy described 
in the preceding section and depicted in Tab.~\ref{Tab_schody}. To train and 
test our model without any bias, we have distilled a subset of the above-mentioned 
160 million states so that each of the hierarchy classes appear with an equal 
probability of $1/5$. This means removing superfluous states belonging to more frequently 
occurring classes and, as a result, the size of the subset of states was reduced 
to about $1/3$, see subsection \ref{App:dataset} and Fig.~\ref{Fig_data} 
in the Appendix. We refer to this procedure as equalization of the dataset.
For all states within this subset, collective measurements outcomes $p_{ij}$ 
were calculated. Vectors of length $n$ composed of the values 
$F = \lbrace p_{ij}\rbrace$ represent the input features for the ANN models. 
In later sections we discuss the performance of the ANN models as function 
of the feature length $n$ as well as on the specific choice of the measurement
configurations $p_{ij}$ leading to the best performance for a given value of $n$.

We have programmed all the ANN models using Pytorch Lightning \cite{Paszke}. 
Our ANN models consist of input and output layers of $n$ and 5 (number of 
classes) nodes respectively with two hidden layers in between each 
possessing 512 fully connected nodes. Each layer is preceded 
by a batch normalization layer to accelerate training \cite{ioffe2015batch}. 
Rectified linear activation function is used on every but the output layer, 
where a Softmax activation function is used instead. Loss is measured by
means of the cross-entropy between the predicted $c'$ and true labels $c$. 
The Adam optimizer \cite{kingma2017adam} with learning rate of $10^{-4}$ is used to update the node 
weights upon training. Pytorch Lightning implementation of the models is available 
as Ancillary files.

The subset of input states was divided with the ratio of 12:3:1 to the training, 
validation and test set respectively. While a maximum number of training epochs 
was set to 4096, in practice the training was halted upon observing no 
improvement on the validation set for 10 consecutive epochs. 
Training started with the mini-batch size of $2^{12}=4096$ and then the models were 
finetuned with the mini-batch size of $2^{18}$ as recommended in Ref.~\cite{Smith2018}. 
Results on the validation set after training has stopped are used to 
establish the best performing model as well as the best suitable composition 
of the input feature vector $F$. This way, we have also established the 
hyper-parameters of our models presented in the preceding paragraph.
Finally, the best performing model is tested on the independent test 
set to assess its true performance.

\begin{figure}
\centering
 \includegraphics[width=.75\columnwidth]{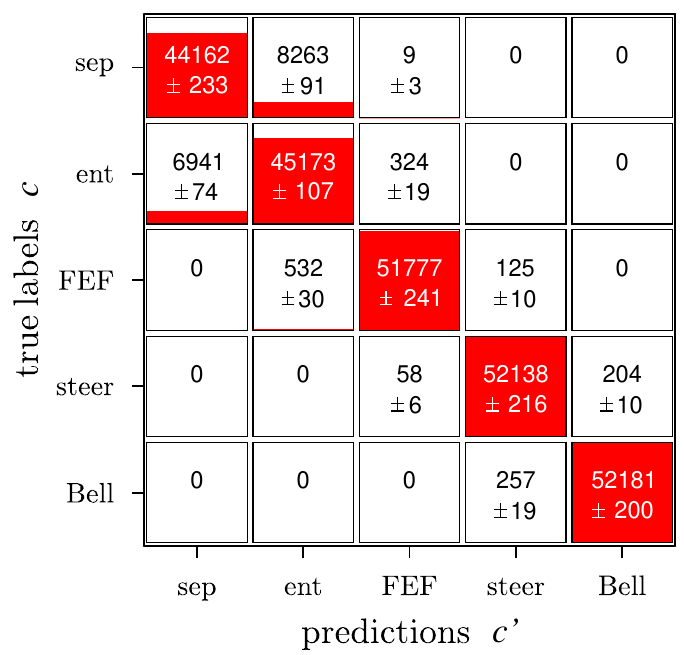}
 \caption{\label{Fig_matrix_Fin10} Confusion matrix for the feature vector $F$ 
           of the full length $n = 10$.}
\end{figure}

Upon testing, the test set was split to 12 subsets in order to estimate statistics
of individual elements of the confusion matrix. Uncertainties typically scale as
square roots of mean values corresponding to Poissonian distribution 
(see Fig.~\ref{Fig_matrix_Fin10}).

\subsection{Benchmarking on full set of measurements}

All correlations except for negativity can be analytically calculated from 
the correlation matrix $R$. We use this fact to asses performance of the ANN 
model when all the necessary information is provided to it, i.e. ANN is 
trained and tested on all $n=10$ measurement configurations $p_{ij}$ 
given by (\ref{EQ_pij}), that suffice to reconstruct the entire $R$ matrix. 
Predictions made by the ANN are accumulated into a confusion matrix CM. 
Element ${cc'}$ of this matrix gives the number of states having the true 
label $c$ while being classified by the ANN as $c'$. The overall accuracy of the model is given by
\begin{equation}
\label{eq:accuracy}
\mathcal{A} = \frac{\mathrm{Tr[CM]}}{\sum[\mathrm{CM}]}.
\end{equation}
Meanwhile recall ${\cal R}$ represents 
the percentage of correctly classified states of a given category
\begin{equation}\label{EQ:Recall}
   {\cal R}_{c} = \mathrm{prob}(c'=c|c). 
\end{equation} 
Precision ${\cal P}$, on the other hand, determines the probability that a state
identified as $c'$ actually belongs to the $(c=c')$ class  
\begin{equation}\label{EQ:Precision}
   {\cal P}_{c'} = \mathrm{prob}(c=c'|c'). 
\end{equation}
This means that recall, and precision is 
the percentage representation of the diagonal element of the confusion matrix 
in a given row and column, respectively. Finally, the $F_1$ score is calculated from 
both of these metrics according to the following formula
which guarantees that it ranges from zero to one
\begin{equation}\label{EQ_F2}
  F_1 = {2 {\cal R P} \over {\cal R+P}}.
\end{equation}

Figure \ref{Fig_matrix_Fin10} shows resulting confusion matrix when the ANN has access to all ten 
elements of the feature vector, $n = 10$. This model was trained on 
a smaller 10M dataset ($10^{7}$ equalized random states) and then finally 
tuned on a bigger 50M dataset ($5\times 10^{7}$ equalized random states).
Resulting $F_1$ score, recall ${\cal R}$ and 
precision ${\cal P}$ for identification of separable, entangled,
FEF, steerable and Bell states are summarized in Tab.~\ref{Tab_Fin10}. 
The performance of the model can be evaluated by specifying the probability 
of the off-diagonal elements, which is in this case $6.38$~\%.
If we do not penalize misclassification
between separable and entangled labels, 
which could not be analytically distinguished from the correlation matrix $R$, 
we are left with only $0.58$~\% of off-diagonal elements. 
The average final accuracy reads $\mathcal{A} = 93.62$~\% or $99.42$~\% when disregarding misclassification between separable and entangled states.
The result of the classification based on the full set of measurements is 
sufficiently precise considering errors that are comparable to analytical 
classification assuming detection shot-noise.
We can, at this point, start reducing the number of measurement configurations
$n$ to make the procedure less resource demanding.  

\begin{table*}
\caption{\label{Tab_Fin10} Recall ${\cal R}$, precision ${\cal P}$ and the $F_1$ 
         score calculated from the confusion matrix in Fig.~\ref{Fig_matrix_Fin10}.} 
%
\centering
\begin{tabular}{c c c c c c}
\hline\hline
 & sep & ent & FEF & steer & Bell \\
\hline
${\cal R}(\%)$ & 84.22 $\pm$ 0.16 & 86.15 $\pm$ 0.13 & 98.74 $\pm$ 0.06 & 99.50 $\pm$ 0.02 & 99.51 $\pm$ 0.04 \\
${\cal P}(\%)$ & 86.42 $\pm$ 0.14 & 83.70 $\pm$ 0.15 & 99.25 $\pm$ 0.04 & 99.27 $\pm$ 0.04 & 99.61 $\pm$ 0.02 \\
$F_1(\%)$       & 85.31 $\pm$ 0.11 & 84.91 $\pm$ 0.10 & 99.00 $\pm$ 0.04 & 99.39 $\pm$ 0.02 & 99.56 $\pm$ 0.02 \\
\hline\hline
\end{tabular}
\end{table*}

\subsection{Finding optimal strategy for measurement reduction 
\label{subsec:reduction}}

\begin{figure}
\centering
 \includegraphics[width=.8\columnwidth]{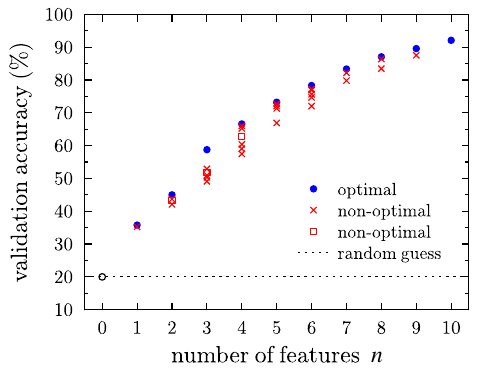}
 \caption{\label{Fig_optimal_10M} 
   Validation accuracy for 10M equalized dataset as a function of the length of 
   feature vector used for training the model. This plot indicates optimality 
   of the selected reduction strategy (blue points). When the reduction of 
   feature elements does not follow the outlined rules, the resulting accuracy 
   is suboptimal (red symbols).}
\end{figure}

First, we have conducted a preliminary testing on a smaller 10M dataset 
to verify optimality of the procedure of measurement reduction (reduction of the length $n$ of the input feature vector). We identify two optimal reduction strategies; one for higher number of projections ($n\geq 5$) and the second for smaller values of $n<5$. Both strategies have their specific physical reasoning based on the inner symmetry of the collective measurements. In case of the first strategy, the ordering of ten elements of the feature vector 
($F = \{1,\ldots,10 \}$) is depicted on the Fig.~\ref{Fig_Features}b). 
The element number 10 is removed as the first then element 9 and so on till 
$n = 5$. Optimal reduction in this range of the values of $n$ is governed by two rules:
(i) It is advantageous to retain the four diagonal feature elements $p_{ii}$ for as 
long as possible, as they contain more information than the six off-diagonal elements.
(ii) If possible, it is favorable not to cumulate canceled elements on a single row or
column.

Continuing with this strategy for $n<5$ would result in suboptimal solutions as depicted  by
red squares in the Fig.~\ref{Fig_optimal_10M}, for $n \in \{2,3,4\}$. 
Therefore, as of $n=4$, we switch to the second strategy motivated by the Collectibility witness \cite{Rudnicki_2012}.
In the next two steps ($n=4$ and $n=3$) we still keep one off-diagonal 
element $p_{ij}$ together with its two corresponding diagonal counterparts $p_{ii}$ 
and $p_{jj}$. In case of $n=4$ we add another diagonal element $p_{kk}$, $k\neq i$ nor $j$. This corresponds to the measurement of the Collectibility 
for a pure state ($n=3$ case) and for a mixed state ($n=4$ case).
It is interesting, that for $n=2$ the optimal strategy is to remove all
diagonal elements keeping two off-diagonal elements $p_{ij}$ and $p_{kl}$, 
where indices $\{i,j,k,l\}$ are exclusively different numbers.    
Finally, for $n=1$ it is again slightly more convenient to keep one diagonal 
element. 
More details are presented in subsection 
\ref{App:schrinking} of the Appendix. 
 
Note that in the case of having no measurements at all ($n = 0$), the accuracy of 
estimation would decrease to 20~\%, i.e. a random guess strategy. 

Correctness of our assertion is verified by numerical testing of several ANN 
models operating on various feature vectors (see Fig.~\ref{Fig_optimal_10M}).
All the models trained with the same symmetry of input feature vectors 
end with similar accuracy with a deviation of 0.7~\%. This contrasts with non-optimal reduction strategies that can render accuracies  more then 10\,\% smaller. Note that there exist 
other established strategies for feature reduction \cite{Rudnicki_2011,Lundberg},
but our approach is favoured because of its clear physical interpretation.
\subsection{Final results obtained following the optimal reduction strategy\label{subsec:optimal}}

\begin{figure}
\centering
 \includegraphics[width=.8\columnwidth]{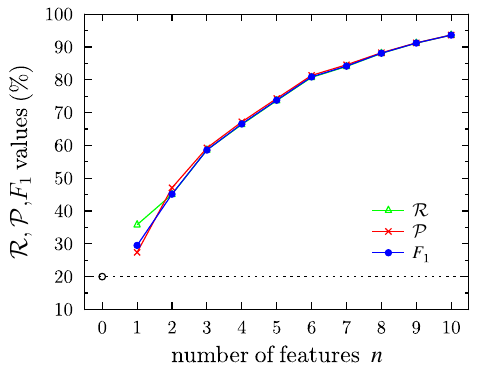}
 \caption{\label{Fig_RPF} Results of the model trained on the 50M dataset 
          using optimal sequences of features of given lengths $n$.
          The points shown represent mean values of recall ${\cal R}$, 
          precision ${\cal P}$ and the $F_1$ score over all categories.}
\end{figure}

\begin{table*}
\centering
\caption{\label{Tab_Fin9} $F_1$ scores for all 5 states' classes and the overall accuracy  $\mathcal{A}$ as functions of feature 
         vector length $n$.} 
%
\begin{tabular}{c c c c c c c}
\hline\hline
\multirow{ 2}{*}{$n$}
  &  \multicolumn{5}{c}{$F_1$ (\%)} & \multirow{ 2}{*}{$\mathcal{A}$ (\%)}   \\ \cline{2-6}
     & sep & ent & FEF & steer & Bell  &                  \\ \hline 
10 & 85.31 $\pm$ 0.11 & 84.91 $\pm$ 0.10 & 99.00 $\pm$ 0.04 & 99.39 $\pm$ 0.02 & 99.56 $\pm$ 0.02 & 93.62 $\pm$ 0.24 \\
9  & 84.35 $\pm$ 0.11 & 80.94 $\pm$ 0.11 & 94.59 $\pm$ 0.07 & 97.56 $\pm$ 0.05 & 98.54 $\pm$ 0.04 & 91.17 $\pm$ 0.25\\
8  & 82.98 $\pm$ 0.11 & 77.01 $\pm$ 0.11 & 88.83 $\pm$ 0.10 & 94.54 $\pm$ 0.07 & 97.13 $\pm$ 0.04 & 88.05 $\pm$ 0.23\\
7  & 80.20 $\pm$ 0.14 & 72.08 $\pm$ 0.12 & 82.09 $\pm$ 0.12 & 91.03 $\pm$ 0.09 & 95.42 $\pm$ 0.07 & 84.05 $\pm$ 0.23\\
6  & 77.37 $\pm$ 0.13 & 67.32 $\pm$ 0.12 & 76.90 $\pm$ 0.13 & 88.51 $\pm$ 0.09 & 94.32 $\pm$ 0.07 & 80.72 $\pm$ 0.22\\
5  & 74.19 $\pm$ 0.14 & 60.10 $\pm$ 0.12 & 64.96 $\pm$ 0.15 & 79.66 $\pm$ 0.11 & 90.00 $\pm$ 0.09 & 73.65 $\pm$ 0.20\\ 
4  & 71.58 $\pm$ 0.13 & 53.91 $\pm$ 0.12 & 54.88 $\pm$ 0.13 & 69.64 $\pm$ 0.12 & 82.73 $\pm$ 0.13 & 66.37 $\pm$ 0.17\\ 
3  & 67.53 $\pm$ 0.15 & 44.67 $\pm$ 0.16 & 45.88 $\pm$ 0.11 & 60.33 $\pm$ 0.12 & 74.50 $\pm$ 0.15 & 58.61 $\pm$ 0.17\\ 
2  & 58.23 $\pm$ 0.15 & 39.86 $\pm$ 0.14 & 35.11 $\pm$ 0.16 & 39.99 $\pm$ 0.20 & 52.56 $\pm$ 0.18 & 45.05 $\pm$ 0.18\\
1  & 52.23 $\pm$ 0.14 & 35.43 $\pm$ 0.11 & 0              & 13.07 $\pm$ 0.09 & 46.85 $\pm$ 0.14  & 35.82 $\pm$ 0.13 \\
\hline\hline
\end{tabular}

\end{table*}

\begin{figure*}
\centering
 \includegraphics[width=2\columnwidth]{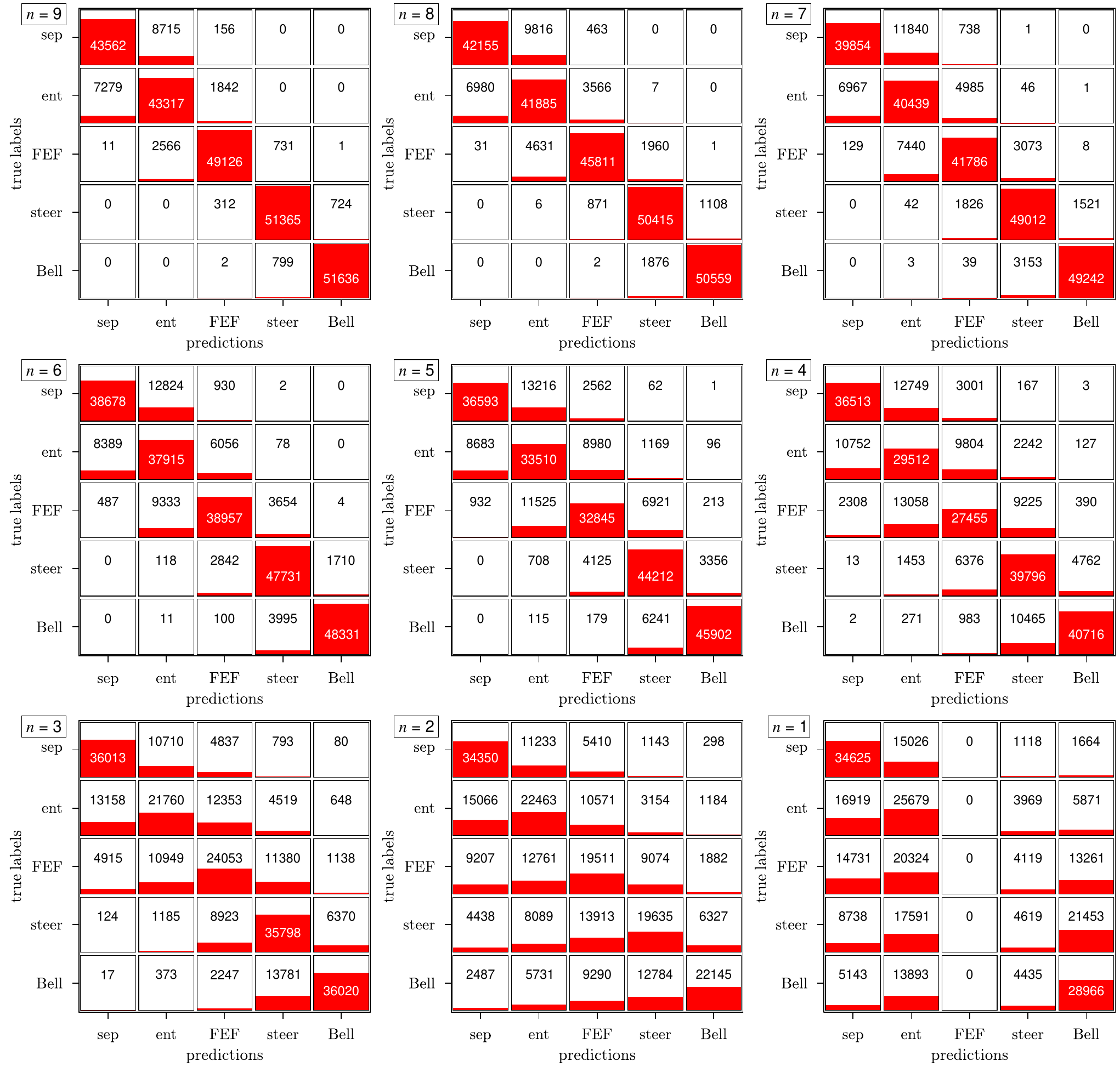}
 \caption{\label{Fig_CM9} Confusion matrices as functions of the feature vector 
          length $n$. Confusion matrix for $n=10$ 
          is shown in Fig.~\ref{Fig_matrix_Fin10}.}
\end{figure*}

Once the optimal feature-reduction procedure has been identified and verified, 
we have proceeded with the final ANN training on the full 50M dataset. 
Figure \ref{Fig_RPF} shows results obtained when reducing the length $n$ of
the feature vector from 10 down to 1. The resulting decrease of the $F_1$ 
scores for all categories of states is also listed in Tab.~\ref{Tab_Fin9}. 
The trend is similar to 
that for the validation set (Fig.~\ref{Fig_optimal_10M}). Only the resulting 
accuracy is slightly better for 50M dataset. 

Figure \ref{Fig_CM9} presents individual confusion matrices for various values of $n$ from which 
the precision and recall are calculated. 
Even though it is impossible to distinguish separable and entangled states by means of analytical calculations, the ANNs achieve this task, albeit with reduced accuracy. 
This figure also shows gradual spreading of the diagonal elements of the confusion matrix when reducing the number of measurements.
Given the equal representation of all classes in the dataset, the sum over any row in the CM yields a constant value. The biggest value is found on the diagonal and it gradually decreases with the decreasing number of measurements $n$. This trend continues down to $n=2$, but for $n=1$ we observe a binarization of the classification splitting the states into two extreme groups: weakly correlated (i.e. separable and entangled) and strongly correlated (i.e. steerable and nonlocal). No state is classified as FEF class. This effect is also witnessed by spreading of precision and recall values for $n=1$ (see Fig.~\ref{Fig_RPF}).

\section{Conclusions} 

This paper has demonstrated the practicality and effectiveness of using ANNs
for classification of quantum correlations in a layout typical for
entanglement-swapping networks. To optimize resource utilization, we
implemented several key measures. Initially, we adopted the minimal basis set,
thereby diminishing the required number of measurement configurations for
analytical classification from 16 to 10. Subsequently, we identified and
numerically tested the optimal measurement reduction strategy, leveraging the
inherent symmetry of the task. This ultimately enabled us to train ANN models
with varying numbers of measurement projections and assess the resulting
precisions, recalls, accuracies, and $F_1$ scores.

To evaluate the performance of the ANN-based classification, we first started
with a model trained using the entire set of 10 measurement configurations and
benchmarked its accuracy against analytical calculations. The ANN model
reached a classification accuracy of 99.42~\% when
allowing for misclassification between separable and entangled states.
It is important to note that the calculation of
negativity, necessary for the proper classification of these two categories,
cannot be performed analytically based on the available measurements. The
achieved accuracy of the ANN models is quite comparable to a perfect
analytical classification assuming inevitable measurement errors in a 
practical environment.
Furthermore, the ANN model demonstrated the capability to label
separable and entangled states with an accuracy of 85.46~\%, a task unfeasible
through conventional analytical calculations. This underscores the benefits of
the ANN classification over the analytical approach.

The adaptability of the ANNs enables the implementation of classification models
using an incomplete set of projections -- a challenge unattainable by
analytical formulae. To observe the diminishing performance of the ANN models,
we reduced the number of measurement configurations from 10 to 1. The performance decline is gradual, with an accuracy of 73.65~\%  achieved with half of the  measurements ($n=5$). 
Finally, an
accuracy of 35.82~\% is obtained with just one single measurement. It is worth noting 
that a random guessing approach, without any measurements, would yield an accuracy
of 20 \%.

Classifying quantum correlations is a crucial challenge in the field of
practical quantum communications networks, particularly when considering the
security implications associated with individual correlations. Consequently,
we believe that our findings make a useful contribution to the ongoing
endeavors aimed at advancing quantum communications toward practical
deployment. Furthermore, the method we have outlined introduces novel research
opportunities in the detection of quantum correlations, addressing, thus, an
intriguing and open fundamental problem in the field.

\acknowledgements

The authors thank CESNET for data management services and J. Cimrman for inspiration. The authors acknowledge support by the project
OP JAC CZ.02.01.01/00/22\_008/0004596
of the Ministry of Education, Youth, and Sports of the Czech Republic and EU.

\section{Appendix}

\subsection{Equalization of the dataset \label{App:dataset}}

\begin{figure}
 \includegraphics[width=\linewidth]{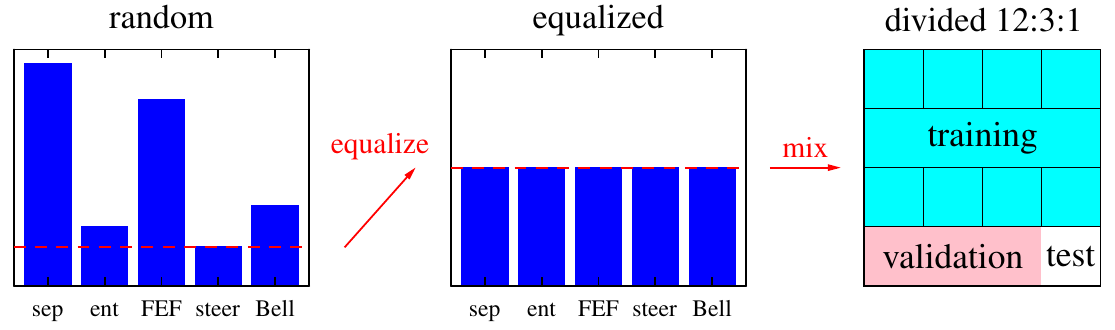}
 \caption{\label{Fig_data} Scheme of the dataset preparation.}
\end{figure}

To train and test the ANN models, we first generate 
a random set of states that are evenly distributed in the space of 
two-qubit states, as was described in Refs. \cite{Pozniak_1998,Maziero2015}.  
Preparation of random datasets is schematically depicted in Fig.~\ref{Fig_data}.
Because individual categories of states are not equivalently represented in the 
raw random dataset, it was necessary to distill an equally balanced (equalized) set. 
The equalization is done by reducing the number of more populated categories 
to the number corresponding to the least populated category, which is steerable
states.
This equalized set is then divided into three parts. The first part is used 
to train the model (the training set). The second part of these 
random equalized states is utilized for validation to stop training in time (prevent overfitting) and compare the performance of various models with different hyperparameters (especially the composition of the input feature vector, see Subsec. \ref{subsec:reduction}). The model that has been the most 
successful in terms of accuracy in recognizing states is chosen for testing 
on the final test set of random equalized states.

\subsection{Minimal required measurements \label{App:minimal}}

\begin{figure}
\centering
 \includegraphics[width=.5\columnwidth]{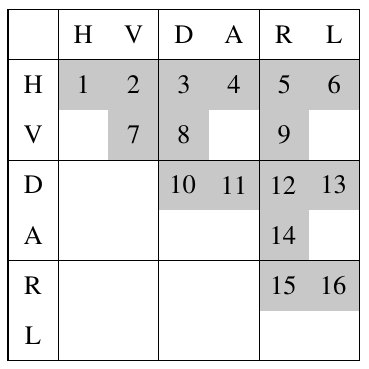}
 \caption{\label{Fig_6x6} Sixteen required von Neumann projections to obtain 
   the correlation matrix $R$ from Pauli measurements. Other not numbered 
   measurements can be derived/deduced using symmetry considerations. 
   Symbols $\{H,V,D,A,R,L\}$ stand for the six eigenstates of the Pauli matrices 
   labelled following the traditional symbolism of polarization optics as 
   horizontal, vertical, diagonal, anti-diagonal, right and left circular,
   respectively.}
\end{figure}

Traditionally, the correlation matrix $R$ is expressed in terms of the Pauli matrices [see Eq. (\ref{eq:R_from_collectives})]. Each Pauli matrix consists of two von~Neumann projectors, hence measuring expectation values of all these matrices on a single qubit requires to perform 6 projective measurements. In case of two qubits, the total number of projective measurements amounts to $6\times 6 = 36$ as illustrated in Fig.~\ref{Fig_6x6}. This number of measurements can be reduced to 16 due to the symmetry of the correlation matrix 
$R_{ij} = R_{ji}$ and also due the completeness relation
\begin{equation}\label{Eq_sigmy}
  \sigma^+_i\sigma^+_j + \sigma^+_i\sigma^-_j + 
  \sigma^-_i\sigma^+_j + \sigma^-_i\sigma^-_j = 1,
\end{equation}
where $\sigma^\pm_i$ are the von Neumann projectors such that 
$\sigma_i = \sigma^+_i - \sigma^-_i$, where $\sigma_i$ is the $i$-th Pauli matrix. Note that the 16th measurement is needed assuming that the overall count rate is not known in advance.

In order to obtain complete information about a single-qubit state, 
it is, however, not necessary to perform 6, but only 4 projection measurements of the minimal basis set $\lbrace\Pi_i\rbrace$ \cite{Rehacek2004}. The directions of the minimal basis projection measurements 
are represented by vertices of a tetrahedron inside the Bloch sphere. Applying this minimal basis set to a two-qubit state requires to perform $4\times 4 = 16$ projective measurements.
All performed projections can be written in a table, where the rows represent 
projections on one qubit, and the columns on the other one.  
Projections performed in the same direction on both qubits are represented by diagonal elements marked grey in the table in Fig.~\ref{Fig_Features}.
Due to the symmetry of $R$, only diagonal and upper off-diagonal elements are unique, this reduces the total number of two-qubit projections to 10.

The correlation matrix $R$ can be easily obtained from the minimal basis projections $\Pi_i$ using the linear relation
$$
  \Pi_i = \sum_j M_{ij} \sigma_j,\quad i,j \in [0,3], \quad \sigma_0 = 1,
$$
where
$$
  M = {1 \over 4}\left(
  \begin{array}{rrrr}
    1 &  s &  s &  s \\
    1 &  s & -s & -s \\
    1 & -s &  s & -s \\
    1 & -s & -s &  s \\
  \end{array}
  \right), \quad s={1 \over \sqrt{3}}.
$$

\subsection{Selected order of the features \label{App:schrinking}}

Here we describe our selection of feature ordering following the 
rules described in sec.~\ref{subsec:reduction}.
The optimality of our reduction strategy can also be deduced from 
Fig.~\ref{Fig_optimal_10M}, where the validation accuracy is plotted as 
a function of the number of the input features.
Firstly, for $n\geq 5$ it is advantageous to retain the diagonal feature elements for as long as 
possible as they contain more information than the off-diagonal elements.
When transitioning from ten measurements to nine, the selection of any
off-diagonal element is arbitrary, so there are six equivalently 
correct possibilities. We remove element $p_{13}$.
After that, for $n=8$, the next removed element is uniquely determined due to symmetry. It must be the 
element $p_{24}$. Removing two elements that intersect on the same row or column 
would be a mistake leading to unnecessary reduction in accuracy.
In the third step ($n=7$), we can again remove any remaining off-diagonal element 
arbitrarily. We selected $p_{34}$, so in the next fourth step ($n=6$) we must remove
$p_{12}$. At this point we have all diagonal elements 
and exactly one off-diagonal measurement remaining in each row and each column.

Subsequently for $n=5$, we remove one of the remaining off-diagonal elements ($p_{23}$) and the 
last remaining off-diagonal element determines the next course of action.

Continuing the reduction for $n<5$, we keep the off-diagonal element $p_{14}$ so we have to keep $p_{11}$ and $p_{44}$ 
when reducing to $n=4$ and $n=3$. In case of $n=4$ one additional diagonal element $p_{33}$ is selected to complement $p_{14}$, $p_{11}$ and $p_{44}$.

For $n=2$ we have to select two off-diagonal elements, i.e. $p_{14}$ and $p_{23}$.
In the last step with only one measurement, it is favourable to employ
a diagonal element: we select $p_{22}$.
In the case of having no measurements at all, the accuracy of estimation 
would be inherently only $1/5 = 20$~\% corresponding to a random guess.

\bibliographystyle{quantum}
\bibliography{citations}


\end{document}